\documentclass{article}
\usepackage{amsmath} 
\usepackage{float}
\usepackage{PRIMEarxiv}
\usepackage{ amssymb }
\usepackage[utf8]{inputenc} 
\usepackage[T1]{fontenc}    
\usepackage{hyperref}       
\usepackage{url}            
\usepackage{booktabs}       
\usepackage{amsfonts}       
\usepackage{nicefrac}       
\usepackage{microtype}      
\usepackage{lipsum}
\usepackage{fancyhdr}       
\usepackage{graphicx}       
\graphicspath{{media/}}     

\title{A  Comparison of Machine Learning Algorithms for Predicting Sea Surface Temperature in the Great Barrier Reef Region
}

\author{
  Dennis Lartey Quayesam \\
  University of Cincinnati \\
  Cincinnati\\
  \texttt{quayesdl@uc.edu} \\
   \And
  Jacob Abugre Akubire \\
  University of Cincinnati  \\
  Cincinnati\\
  \texttt{akubirja@uc.edu} \\
   \And
    Oliveira Ann Gyamfua Darkwah \\
  University of Cincinnati  \\
  Cincinnati\\
  \texttt{darkwaoa@uc.edu} \\
}

\begin{document}
\maketitle

\begin{abstract}
Predicting Sea Surface Temperature (SST) in the Great Barrier Reef (GBR) region is crucial for the effective management of its fragile ecosystems. This study provides a rigorous comparative analysis of several machine learning techniques to identify the most effective method for SST prediction in this area. We evaluate the performance of ridge regression, Least Absolute Shrinkage and Selection Operator (LASSO), Random Forest, and Extreme Gradient Boosting (XGBoost) algorithms. Our results reveal that while LASSO and ridge regression perform well, Random Forest and XGBoost significantly outperform them in terms of predictive accuracy, as evidenced by lower Mean Squared Error (MSE), Mean Absolute Error (MAE), and Root Mean Squared Prediction Error (RMSPE). Additionally, XGBoost demonstrated superior performance in minimizing Kullback-Leibler Divergence (KLD), indicating a closer alignment of predicted probability distributions with actual observations. These findings highlight the efficacy of using ensemble methods, particularly XGBoost, for predicting sea surface temperatures, making them valuable tools for climatological and environmental modeling.
\end{abstract}

\keywords{ Random Forest \and XGBoost \and Sea Surface Temperature}

\section{Introduction}

Sea Surface Temperature (SST) is a fundamental parameter in oceanography and ocean technology, playing a pivotal role in various aspects of marine science and industry. It is crucial for comprehending and predicting weather and climate dynamics, as well as planning diverse offshore activities \cite{patil2016prediction}. Extreme high sea surface temperatures have emerged as a significant factor contributing to global warming over the past three decades \cite{yang2017cfcc}. Observations of SST are essential for understanding the complex interactions between the ocean and the global climate \cite{usharani2023ilf}. 

Predicting Sea Surface Temperature (SST) is essential for managing marine ecosystems, particularly in regions like the Great Barrier Reef (GBR), where delicate environmental balances are vital for biodiversity and ecological health. The Great Barrier Reef, situated along the coast of Queensland in northeastern Australia, is the largest coral reef ecosystem in the world. However, despite its ecological and economic importance, the Great Barrier Reef faces a myriad of threats that jeopardize its long-term survival. According to \cite{daley2014great}, the Great Barrier Reef faces escalating challenges stemming from tourism, fishing, pollution, and climate change . This iconic and invaluable marine ecosystem is under unprecedented threat, primarily due to climate change and human activities. 
Sea surface temperature emerges as a critical parameter for assessing the reef's health, as it influences coral growth, reproduction, and resilience to stressors. Climate change-induced phenomena, such as rising sea temperatures and ocean acidification, have led to widespread coral bleaching events. Furthermore, the reef is subject to pollution from agricultural runoff, coastal development, overfishing, and other human activities, exacerbating its already fragile state. 

As climate change intensifies and anthropogenic pressures mount, accurate prediction of sea surface temperature (SST) in the Great Barrier Reef (GBR) region has become increasingly vital. Understanding SST is crucial for deciphering ecological processes, identifying emerging threats, and guiding effective conservation efforts. The ability to predict SST with high precision allows scientists to anticipate and mitigate the impacts of coral bleaching, monitor changes in marine biodiversity, and develop adaptive management strategies. Furthermore, reliable SST predictions contribute to the broader goal of preserving the ecological integrity of the GBR, ensuring that this natural wonder continues to thrive amidst growing environmental challenges.

Against this backdrop, this study seeks to evaluate the performance of penalized linear regression models (Lasso and Ridge) and tree-based regression models (Random Forest and XGBoost) in predicting sea surface temperature in the Great Barrier Reef region. The application of machine learning techniques for this purpose has gained traction due to their ability to model the complex, nonlinear relationships inherent in environmental data. By comparing their predictive accuracy and interpretability, we aim to identify the most suitable technique for generating reliable SST forecasts, thereby contributing to the broader understanding of reef dynamics and informing evidence-based conservation strategies.

In essence, this study endeavors to harness the power of machine learning to address pressing environmental challenges and safeguard the ecological integrity of the Great Barrier Reef, an endeavor critical for the preservation of one of the world's most iconic natural treasures. Traditional statistical methods have been employed for SST prediction, but achieving accurate prediction of SST is essential for understanding the reef's dynamics, identifying potential threats, and formulating effective conservation strategies.  Various studies have explored different regression techniques for SST modeling\cite{miftahuddin2021modeling}.  

According to Shalabh (2020)\cite{shalabh2020statistical}, linear regression, a widely used statistical technique, is foundational in environmental prediction tasks due to its simplicity and interpretability. A study conducted by Miftahuddin \cite{miftahuddin2016fundamental} utilized linear regression models to analyze the relationship between SST and climate parameters, highlighting the significance of time covariates in model fitting. Hsieh (2020)\cite{hsieh2020improving} states that while linear regression is effective in modeling SST variations in the GBR region using predictor variables such as sea surface salinity and atmospheric indices, it may struggle with the complex nonlinear relationships inherent in SST dynamics, thereby impacting its predictive accuracy.  Petrenko et al. \cite{petrenko2015suppressing} discussed the use of multichannel regression algorithms in retrieving SST from satellite observations, emphasizing noise suppression techniques to improve SST retrieval accuracy. Additionally, He, Zhang, and Wang \cite{he2020development} emphasized the importance of regional algorithms in estimating SST using satellite remote sensing in the western North Pacific. Wu and Zhang \cite{wu2023enhancing} introduced a symbolic regression (SR) approach to enhance the predictive capabilities of turbulence models. 

In recent years, machine learning techniques have emerged as potent tools for predicting environmental variables, holding the promise of surpassing traditional statistical methods. The rise of machine learning brings exciting new ways to improve predictions and discover hidden patterns in complex data. Through this study, we aim to advance understanding of predictive modeling techniques and inform evidence-based management strategies for preserving the ecological integrity of the Great Barrier Reef. The study proceeds with a discussion of the methods used in section 2, analysis  and discussion of results in section 3 and conclusion in section 4

\section{Methodology}

\subsubsection{LASSO}
The Least Absolute Shrinkage and Selection Operator (LASSO) method represents a pivotal advancement in regression analysis, combining variable selection with regularization to enhance model robustness and interpretability. This technique modifies regression estimation by introducing a penalty term proportional to the absolute size of the coefficients, helping to address multicollinearity and highlight significant predictors. Notably, LASSO can compress the coefficients of non-essential variables to zero, effectively omitting them from the model. This attribute is particularly beneficial in scenarios characterized by multicollinearity, as it enhances both the interpretability and accuracy of model predictions. The efficacy of LASSO in this context has been rigorously documented in seminal works, such as Tibshirani (1996)\cite{tibshirani1996regression}. Furthermore, LASSO proves invaluable in high-dimensional settings, as outlined by Meinshausen et al.\cite{meinshausen2006variable}, where it efficiently estimates sparse coefficients, even when the number of potential predictors surpasses the available sample size. It is formulated in \cite{hastie2009elements}as follows

\begin{equation}
\hat{\beta}^{lasso} = \arg \min_{\beta} \left\{ \frac{1}{2} \sum_{i=1}^{N} (Y_i - \beta_{0}- \sum_{j=1}^{p}X_ij\beta_j)^2 + \lambda \sum_{j=1}^{p} |\beta_j| \right\}
\end{equation}

where $\beta$ 's are the coefficients and $\lambda$ is the tuning parameter.
where $\lambda$ is the tuning parameter that controls the strength of the penalty, $Y_{i}$ are the observed values, $X_ij$ are the predictor variables, $\beta$'s are the parameters to be estimated, and $p$ is the number of predictors.

\subsubsection{Ridge Regression}
Ridge regression represents a sophisticated variant of linear regression specifically designed to counteract the challenges posed by collinearity among predictors. This method enhances the stability and predictive accuracy of linear regression models by integrating a ridge parameter, which optimally balances bias and variance. The inclusion of $L_2$
regularization in ridge regression reduces the model's complexity and effectively addresses multicollinearity among independent variables. This regularization technique penalizes the magnitude of the coefficients, thereby constraining them to smaller values and ensuring a more generalized model formulation\cite{hilt1977ridge}. It is formulated in \cite{hastie2009elements} as follows:
\begin{equation}
\hat{\beta}^{ridge} = \arg \min_{\beta} \left\{ \sum_{i=1}^{N} (Y_i - \beta_{0}- \sum_{j=1}^{p}X_ij\beta_j)^2 + \lambda \sum_{j=1}^{p} \beta_j^2 \right\}
\end{equation}

\subsubsection{Random Forest}
Random Forest is a powerful machine learning method that combines multiple decision trees through bagging \cite{popuri2022approximation}. It is widely used for classification and regression tasks due to its ability to reduce variance and overfitting, making it robust for various applications \cite{hashemi2024robust}. Random Forest models can be optimized by adjusting hyperparameters such as the number of estimators, features, depth, split, and leaf nodes. For a detailed formulation of Random Forest, see \cite{breiman2001random}. The Random Forest prediction is obtained by averaging the predictions from multiple decision trees:

\begin{equation}
 \hat{Y} = \frac{1}{N} \sum_{i=1}^{N} \hat{Y}_i   
\end{equation}  

where:
\begin{itemize}
  \item \( \hat{Y} \) is the final prediction.
  \item \( N \) is the number of decision trees in the forest.
  \item \( \hat{Y}_i \) is the prediction from the \( i \)-th decision tree.
\end{itemize} 
 
\subsubsection{XGBoost}
Extreme Gradient Boosting (XGBoost) is an advanced ensemble learning method that has gained widespread acclaim within the machine learning community for its robust performance across a variety of tasks. This method operates by aggregating the predictions from multiple "weak" models to form a singular, more accurate, and powerful prediction, as detailed in Chen and Guestrin (2015)\cite{chen2015xgboost}. XGBoost distinguishes itself by its scalability and efficiency in processing large datasets, attributes that contribute to its extensive adoption in both classification and regression challenges. The strength of XGBoost lies in its sophisticated algorithmic structure, which systematically enhances weak models through an optimization framework specifically tailored for boosted trees. The XGBoost model is formulated as described in \cite{chen2015xgboost},
\begin{equation}
    \hat{Y_{i}}=\phi(X_{i})=\sum_{k=1}^k f_{k}(X_{i}) \;,\;\;  f_{k}\in \mathcal{F}
\end{equation}
In XGBoost, we consider $k$ trees, where $\mathcal{F} = {f(x) = w_{q(X)}}$ represents a set of Classification and Regression Trees (CART). Here, $q$ denotes the mapping function of each independent decision-tree structure to its corresponding leaves, and $w_{q(X)}$ is the weight assigned to the leaf that an input $X$ maps to. The ensemble set $\mathcal{F}$ is optimized by minimizing the objective function
\begin{equation}
  \mathcal{F}_{\phi}=  \sum_{i} \iota(\hat{Y_{i}},Y_{i}) + \sum_{k} \omega (f_{k})
\end{equation}
with $$ \omega(f)=\frac{1}{2}\lambda ||w|| ^2$$

\subsection{Model Evaluation}
The performance of each model is meticulously evaluated using a comprehensive suite of statistical metrics applied to the testing data. These metrics include Mean Squared Error (MSE), Mean Absolute Error (MAE), Root Mean Square Percentage Error (RMSPE), and Kullback-Leibler Divergence (KLD). This robust assessment framework ensures a thorough analysis of model accuracy and predictive reliability across varying statistical dimensions.

The MSE measures the average squared difference between the observed and predicted values. Lower values indicate better model performance. The MSE is computed as follows:
\[ \text{MSE} = \frac{1}{n} \sum_{i=1}^{n} (Y_i - \hat{Y}_i)^2 \]

 where $ n $ is the number of observations, $Y_i$ is observed value and 
   $\hat{Y}_i$ is the predicted value.

The MAE represents the average absolute difference between the observed and predicted values. Similar to MSE, lower values signify better accuracy. The MAE is computed as follows:
\[ \text{MAE} = \frac{1}{n} \sum_{i=1}^{n} |Y_i - \hat{Y}_i| \]

where $ n $ is the number of observations, $Y_i$ is observed value and 
   $\hat{Y}_i$ is the predicted value

The RMSPE reflects the square root of the mean squared prediction errors, providing a measure that is on the same scale as the data. The RMSPE formula is given by:
\[ \text{RMSPE} = \sqrt{ \frac{1}{n} \sum_{i=1}^{n} \left( \frac{Y_i - \hat{Y}_i}{Y_i} \right)^2 } \]

The KLD measures the divergence between the predicted probability distribution and the true distribution. A lower KLD value indicates that the predicted distribution is closer to the true distribution. The formula for KLD is given by
\[ \text{KLD}(P \| Q) = \sum_{i=1}^{n} P(X_i) \log \left( \frac{P(X_i)}{Q(x_i)} \right) \]

where $ P(X_{i}) $ represent the true probability distribution,
    $Q(X_{i})$: Predicted probability distribution 
\newpage
\section{Discussion of  results}
This study utilizes comprehensive datasets to analyze Sea Surface Temperatures (SST) in the Great Barrier Reef (GBR) region, spanning from 2003 to 2020. Below, we outline the key variables of the dataset:
\begin{itemize}
\item  Latitude (lat) and Longitude (lon): These variables represent the geographical coordinates of the data points. 
\item Higher-order Interactions: Interaction terms such as lat.n1, lat.n2, lon.n1, and lon.n2  account for non-linear spatial effects and complex interactions between geographical coordinates.
\item General Circulation Models (GCMs) Variables:
These variables ( GCM.n1, GCM.n2, GCM.n3, GCM.n4) are derived from global climate system, under various climatic scenarious.
\end{itemize}
The dataset is meticulously partitioned into training and testing sets.
\subsection{Correlation Analysis}
A correlation matrix was computed to examine the linear relationships between features. The figure below shows the correlation between the SST and the other variables.
\begin{figure}[h]
  \centering
  \includegraphics[width=0.6\linewidth]{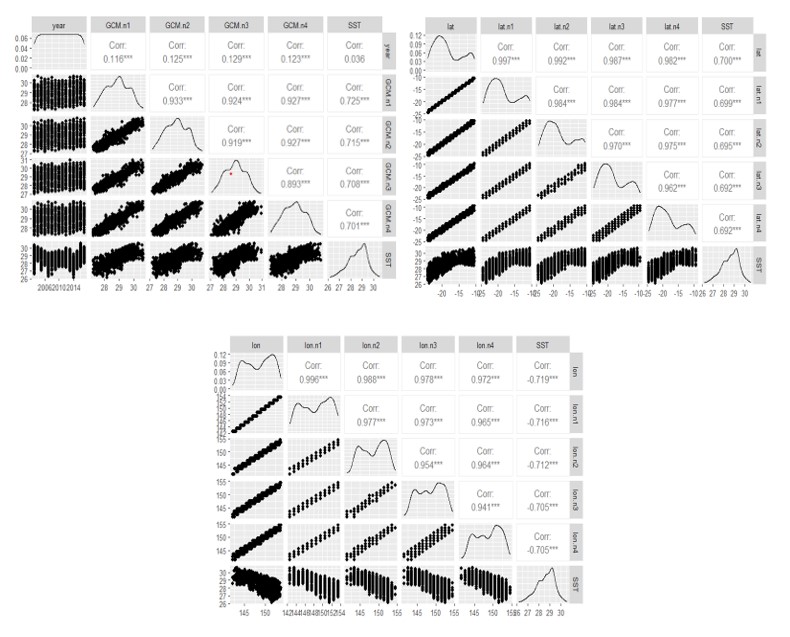} 
  \caption{Correlation between SST and other variables}
  \label{fig:fig1}
\end{figure}

Figure 1 presents correlation plots illustrating the relationships between key predictor variables and sea surface temperatures (SST), highlighting significant correlations among Global Climate Models (GCMs), spatial variables (latitude and longitude), and SST. The top-left panel shows strong positive correlations between GCM variables (GCM.n1, GCM.n2, GCM.n3, GCM.n4) and SST, with correlation values as high as 0.933. The top-right panel indicates a robust positive correlation between latitude variables (lat, lat.n1, lat.n2, lat.n3, lat.n4) and SST, with values ranging from 0.692 to 0.700. Conversely, the bottom panel depicts strong negative correlations between longitude variables (lon, lon.n1, lon.n2, lon.n3, lon.n4) and SST, with values such as -0.716 for lon.n3 and -0.712 for lon.n2. Additionally, the correlation plots reveal high intercorrelations among the independent variables themselves.

\newpage

\begin{figure}[h]
  \centering
  \includegraphics[width=0.6\linewidth]{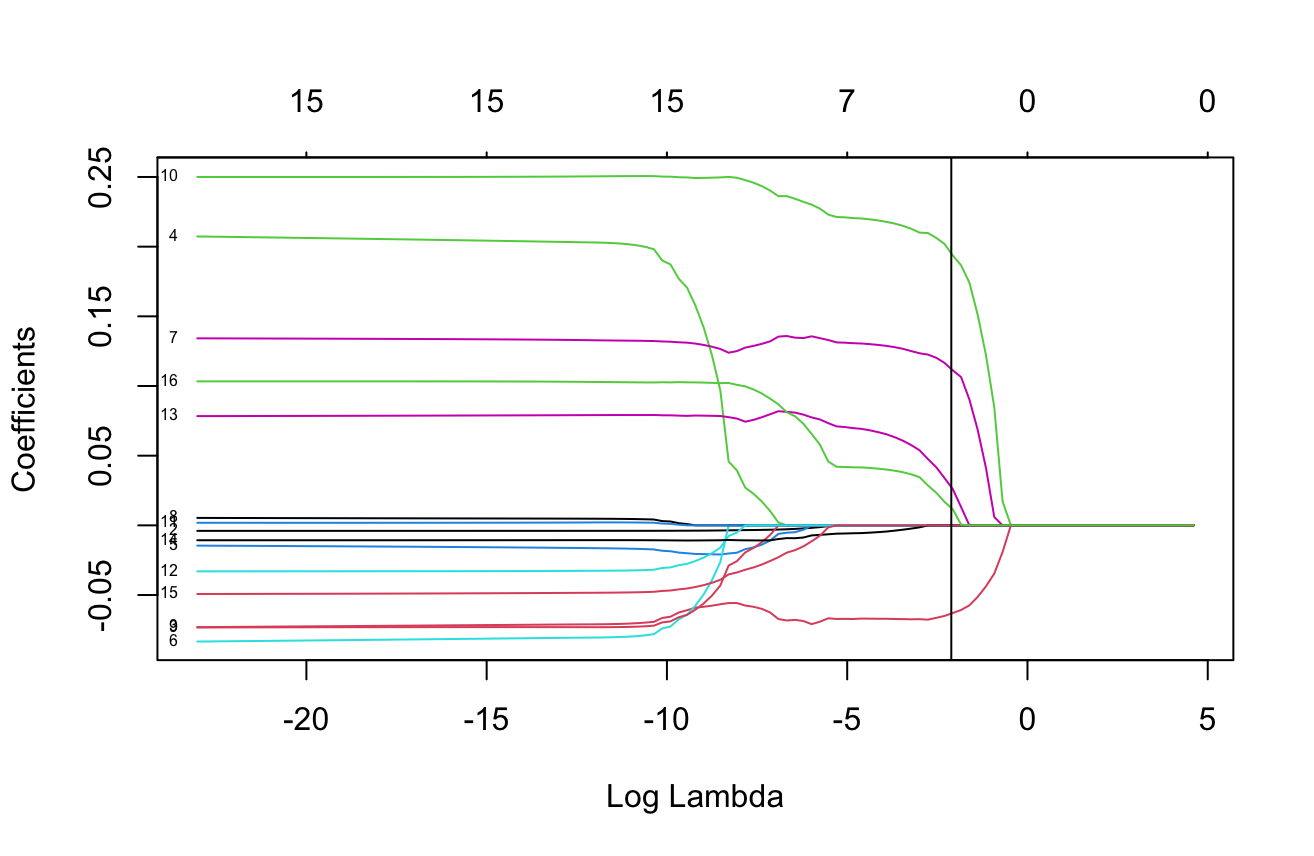} 
  \caption{Trajectories of Predictor Coefficients in Lasso Regression as a Function of Logarithmic Lambda for Sea Surface Temperature}
  \label{fig:fig2}
\end{figure}

Figure 2 illustrates the profile plot of coefficients for various predictors in a Lasso regression model.  Each line represents a different predictor, with the path showing how its coefficient changes as the logarithm of the penalty parameter $\lambda$ increases. As as we move from a log $\lambda$ of -20 to 5, several coefficients shrink significantly toward zero, indicating their relatively lesser importance or contribution to the model. In contrast, a few coefficients remain non-zero even at higher $\lambda$ values, suggesting these variables are key drivers in predicting sea surface temperatures.

Five predictors exhibit significant non-zero values across a wide range of $\lambda$, highlighting their strong influence on the model. Several coefficients stabilize at zero early as $\lambda$ increases, particularly those represented in blue, which become negligible, indicating a minimal or redundant impact on model predictions at higher penalty values.

The plot below is essential for understanding the predictive performance behavior of the Lasso model under varying degrees of regularization and for selecting an optimal $\lambda$ that minimizes prediction error.

\begin{figure}[h]
  \centering
  \includegraphics[width=0.6\linewidth]{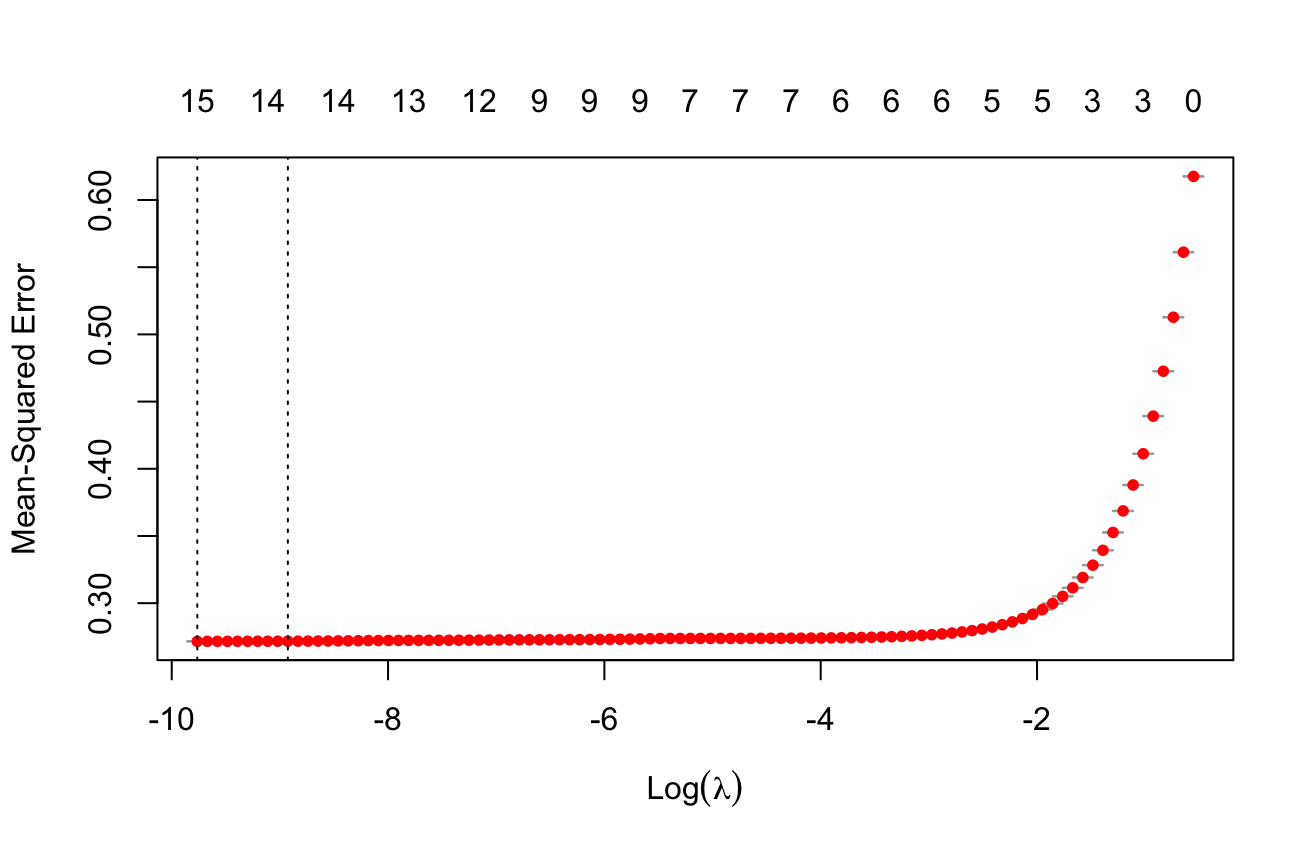} 
  \caption{Impact of Regularization Parameter on Model Performance in Lasso Regression}
  \label{fig:fig3}
\end{figure}

Figure 3 above illustrates the dependency of the mean squared error (MSE) on the logarithm of the regularization parameter $\lambda$ in a Lasso regression model.  As $log(\lambda)$ increases from -10 to 0, the MSE remains relatively stable and low, indicating that the model is able to retain its predictive accuracy despite the increasing regularization. This plateau suggests that a wide range of $\lambda$ values around this region could be optimal for balancing model complexity and prediction error.

However, as $log(\lambda)$ progresses past -4 towards 0, there is a sharp upward curve in the MSE, illustrating a significant deterioration in model performance. This increase in MSE at higher $\lambda$ values is indicative of over-regularization, where too many predictor coefficients are driven to zero, leading to underfitting and a loss of important information.

The dotted vertical lines at $log(\lambda)$ values of approximately -8 and -6 might represent thresholds beyond which the increase in $\lambda$ no longer contributes to a decrease in MSE. These lines could be suggesting candidate values for $\lambda$ that minimize the MSE, thereby providing a practical guide for parameter selection in Lasso regression models.

\begin{figure}[h]
  \centering
  \includegraphics[width=0.6\linewidth]{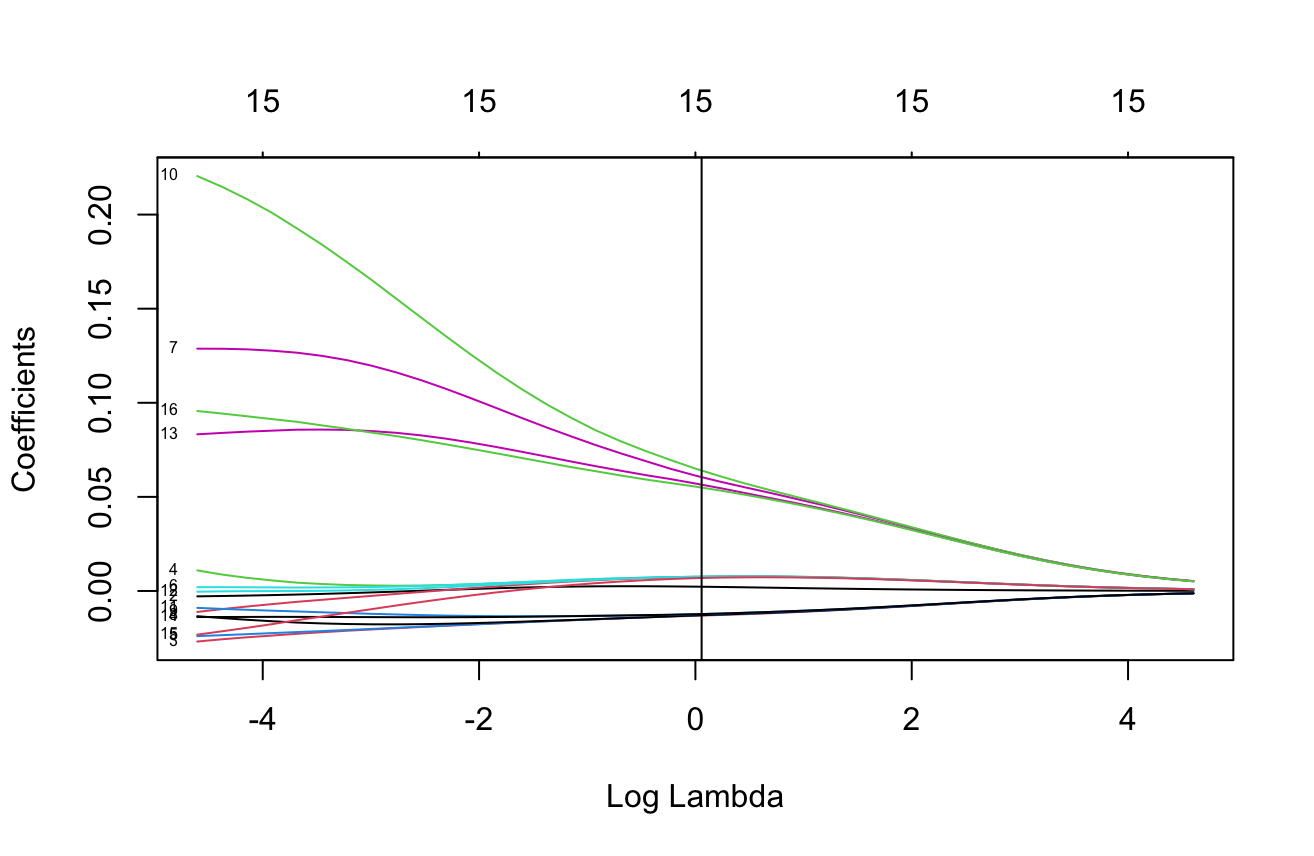} 
  \caption{Evolution of Coefficients in Ridge Regression with Varying Logarithmic Lambda}
  \label{fig:fig4}
\end{figure}

Figure 4 shows the trajectories of coefficients for several predictors in a Ridge regression model, focusing on their behavior as the regularization parameter $\lambda$ varies. 

As $\lambda$ increases, all coefficients gradually shrink towards zero.  Initially, when $log(\lambda)$ is low (e.g., between -4 and 0), the coefficients of four important predictors are significantly different from zero, indicating that the model retains much of the influence from those predictors. As $log(\lambda)$ increases beyond 0, a rapid decline in coefficient values is observed, reflecting strong regularization effects where the model significantly reduces the influence of less critical predictors. Yet, few predictors  show a more gradual decline compared to others, suggesting their stronger or more consistent contribution to the model across varying degrees of regularization. These predictors are potentially more significant for predicting the target variable, hence their coefficients are more resistant to shrinkage.

\begin{figure}[h]
  \centering
  \includegraphics[width=0.6\linewidth]{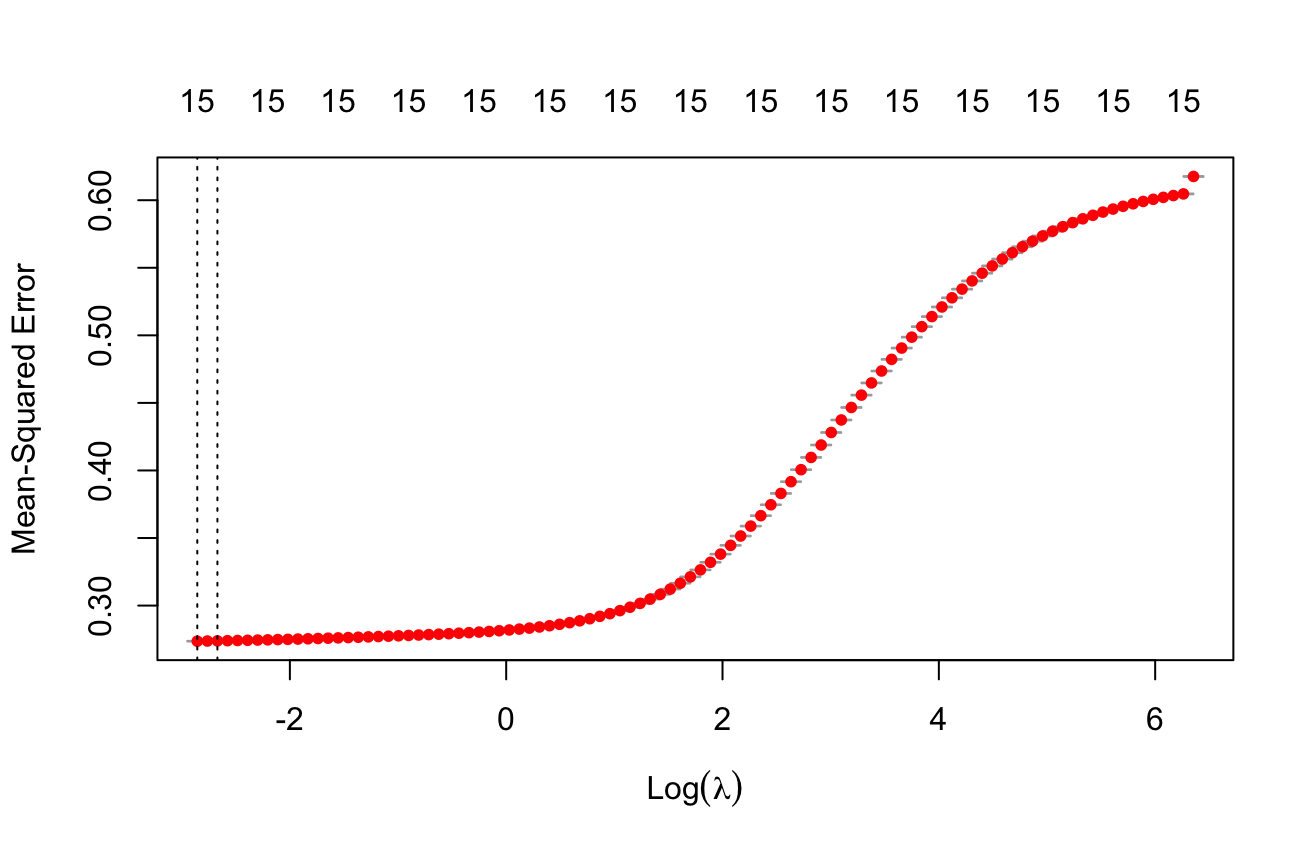} 
  \caption{Mean Squared Error as a Function of Logarithmic Lambda in Ridge Regression}
  \label{fig:fig5}
\end{figure}

 Figure 5 above depicts the mean squared error (MSE) of a Ridge regression model as a function of the logarithm of the regularization parameter $\lambda$. For values of $Log(\lambda)$ less than zero, the MSE remains relatively constant and minimal, indicating that the model with lower regularization strengths performs well and is capable of capturing the underlying pattern without substantial bias. The dotted vertical line around  represent the transition point beyond which increasing the regularization parameter begins to adversely affect the model’s performance. This point is potentially considered as an optimal balance between bias and variance. As $Log(\lambda)$ increases beyond 0, there is a clear upward trend in MSE, which becomes pronounced as $ Log(\lambda)$ approaches 6. This trend indicates that excessive regularization introduces high bias, leading to underfitting where the model fails to capture the essential relationships in the data, thereby increasing prediction errors.

\begin{figure}[h]
  \centering
  \includegraphics[width=0.6\linewidth]{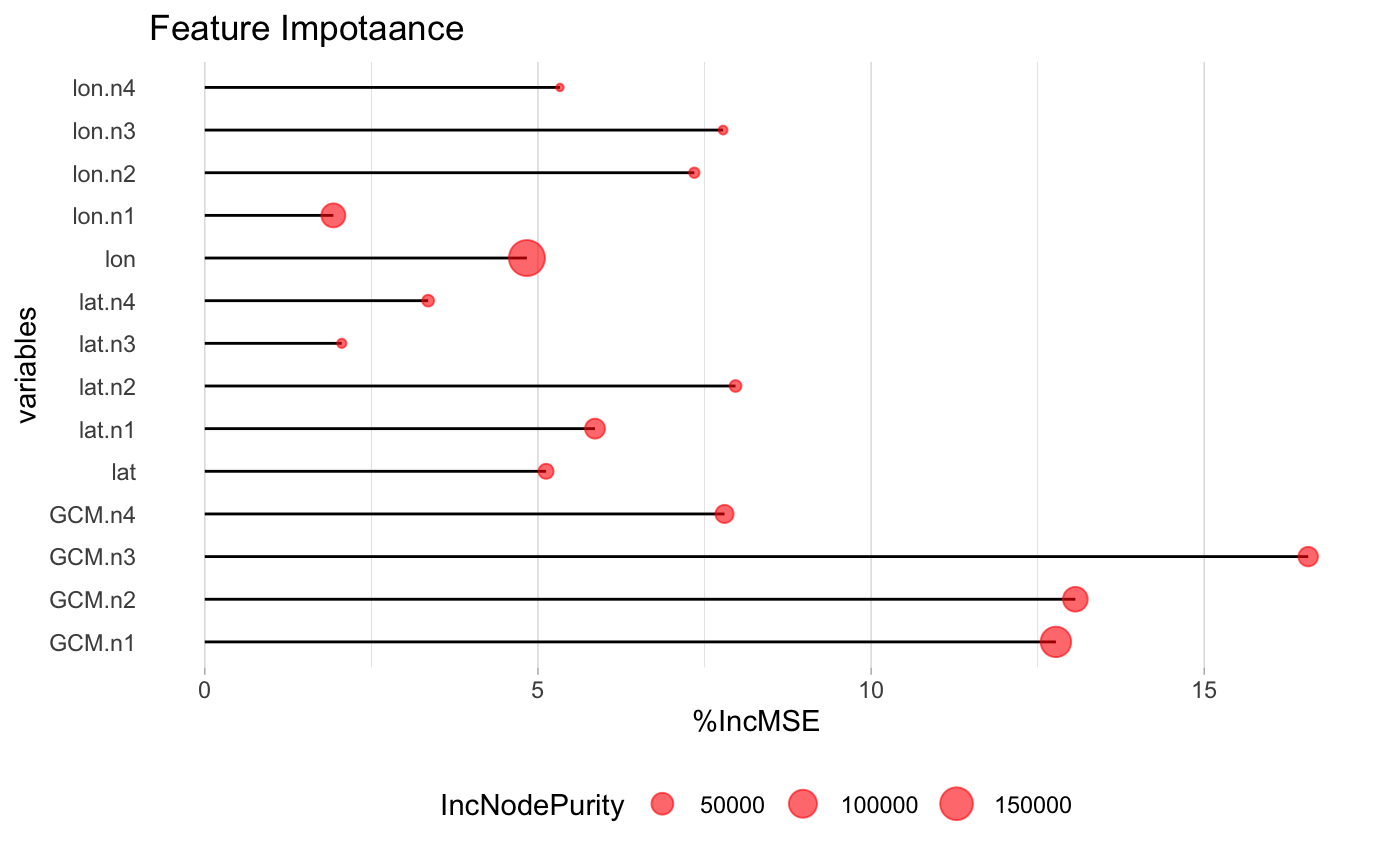} 
  \caption{ Feature Importance Analysis in Random Forest Model for Sea Surface Temperature Prediction}
  \label{fig:fig 6}
\end{figure}

 Figure 6 presents the importance of various features in the Random Forest model used for predicting sea surface temperatures. The x-axis shows two metrics for feature importance: the percentage increase in mean squared error (\%IncMSE) and the increase in node purity (IncNodePurity). The y-axis lists the predictor variables, which include latitude (lat), longitude (lon), and variables from the Global Climate Model (GCM). The \%IncMSE metric measures the increase in the mean squared error of the model predictions when a particular feature is permuted. Higher values indicate that the feature is crucial for accurate predictions. The IncNodePurity metric represents the total decrease in node impurity (measured by the Gini impurity or another criterion) attributed to each feature, averaged over all the trees in the forest. Larger values indicate a greater importance of the feature in improving the model's accuracy.

The Longitude (lon) and Latitude (lat) are among the most important predictors, with longitude (lon) and its first-order interaction term (lon.n1) showing high \%IncMSE and IncNodePurity values. This highlights their significant role in predicting sea surface temperature variations.
The GCM variables, particularly GCM.n1, GCM.n2, and GCM.n4, also exhibit substantial importance, reflecting the influence of global climate patterns on sea surface temperatures. Interaction terms such as lon.n1, lat.n1, and others suggest that interactions between latitude and longitude variables play a crucial role in the model, capturing complex spatial dependencies.

The visual representation emphasizes that spatial variables (latitude and longitude) and their interactions are key drivers in the model, underlining the importance of spatial resolution in sea surface temperature modeling. Additionally, the significant contribution of GCM variables corroborates the influence of broader climatic factors, suggesting that integrating local spatial data with global climate indicators enhances the predictive power of the model.

\begin{figure}[H]
  \centering
  \includegraphics[width=0.6\linewidth]{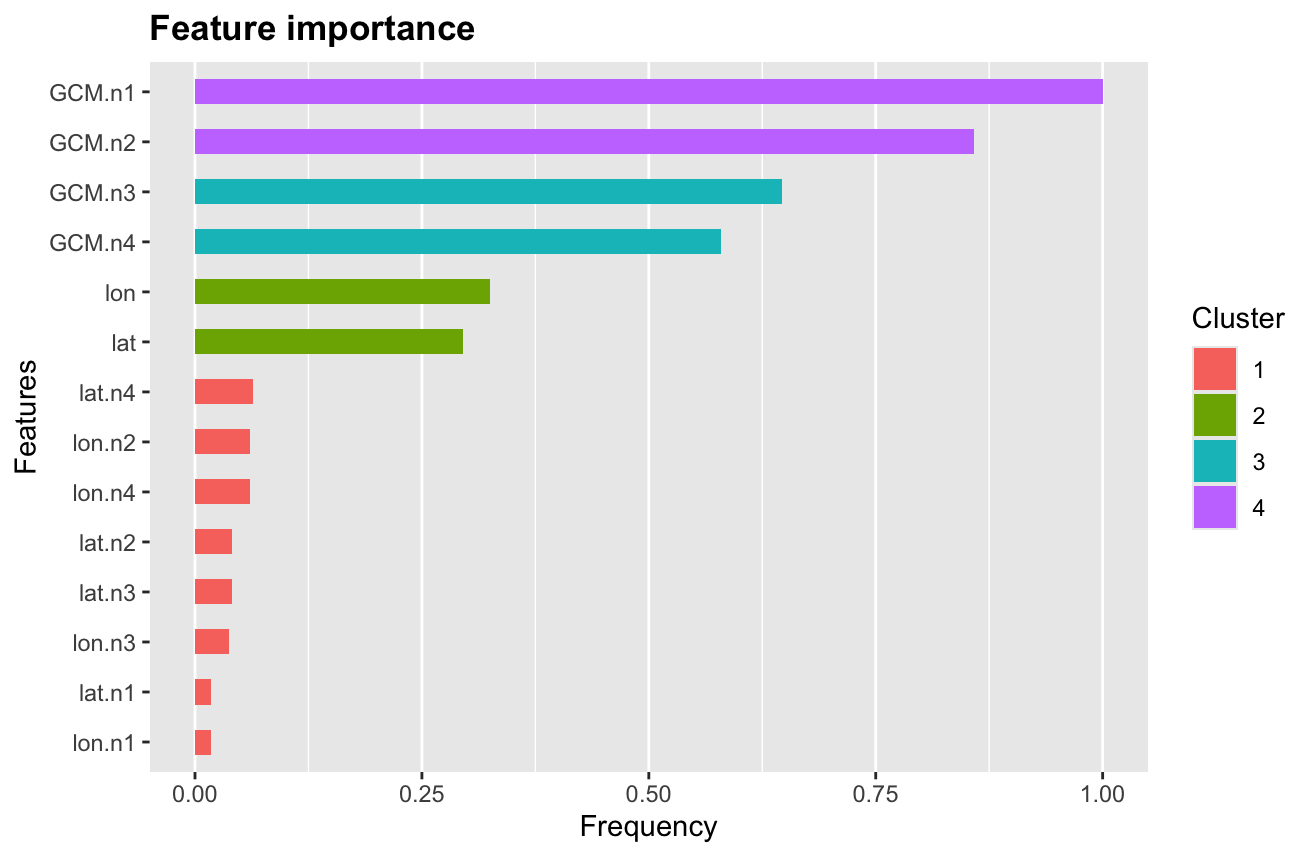} 
  \caption{ Feature Importance Analysis in XGBoost Model for Predicting Sea Surface Temperature}
  \label{fig:fig 7}
\end{figure}

Figure 7 displays the importance of different features in the XGBoost model used to predict sea surface temperatures. The x-axis shows the frequency or importance score of each feature, while the y-axis lists the features. Features are grouped into clusters, as indicated by the color coding. The frequency metric represents how often a feature was used in the trees of the XGBoost model. A higher frequency indicates a more important feature, meaning it was more frequently used for splits in the decision trees, thereby contributing more to the model’s predictive power.
Key Observations:

The General Circulation Models (GCMs) variables (GCM.n1, GCM.n2, GCM.n3, GCM.n4) are the most important features, as indicated by their high frequency scores. This suggests that these variables play a crucial role in predicting sea surface temperatures. The Longitude (lon) and latitude (lat) also exhibit significant importance, highlighting their relevance in the model. Interaction terms such as lat.n4 and lon.n2 show moderate importance, suggesting that while they contribute to the model, their impact is less significant compared to primary spatial variables and GCM components.

\begin{figure}[H]
  \centering
  \includegraphics[width=0.6\linewidth]{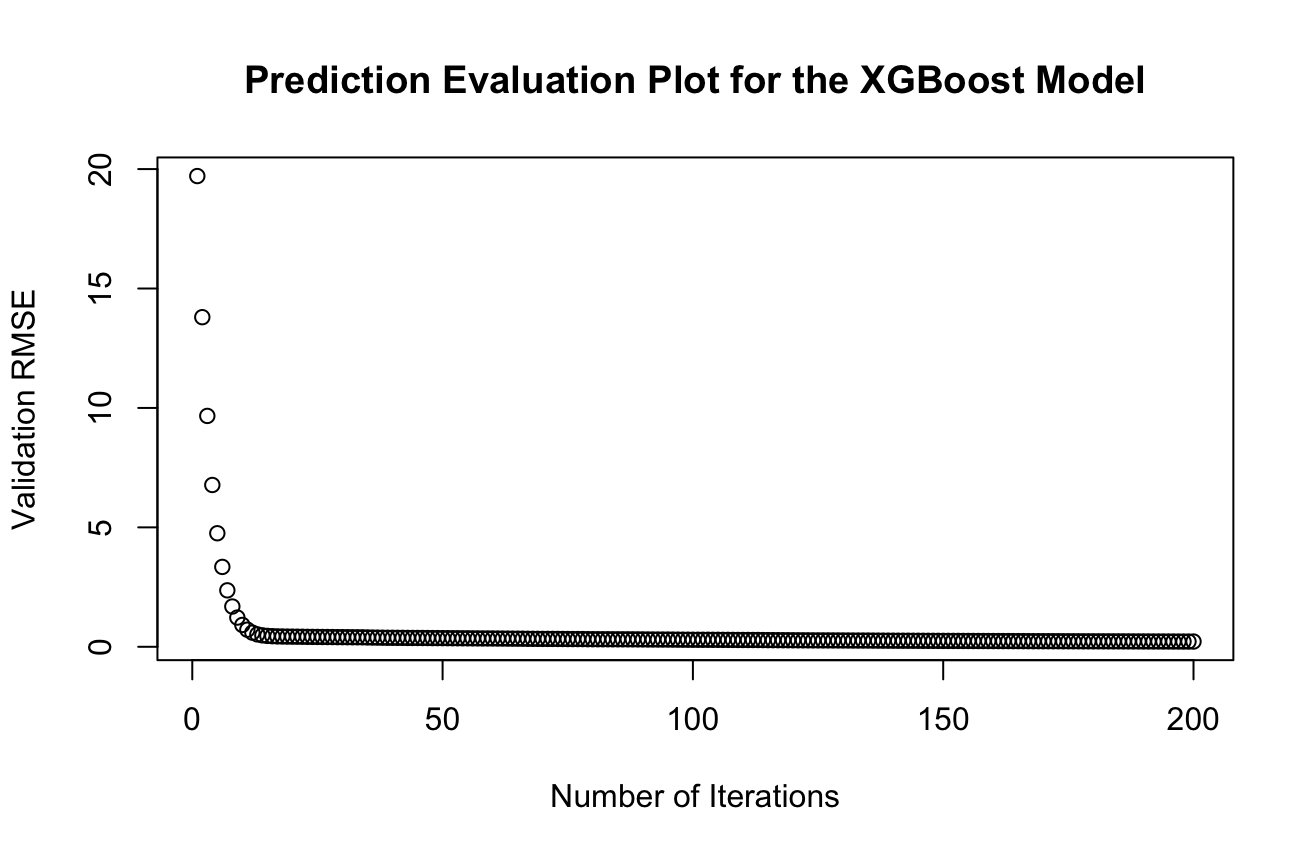} 
  \caption{ Validation RMSE Convergence over Iterations in XGBoost Model}
  \label{fig:fig 8}
\end{figure}

Figure 8 above presents the validation root mean squared error (RMSE) for an XGBoost model as a function of the number of iterations. At the beginning of the training process there is a steep decline in the validation RMSE. This indicates that the model quickly learns the underlying patterns in the data and significantly improves its predictive accuracy. After the initial rapid decrease, the RMSE stabilizes and shows minimal change with further iterations. This plateau suggests that the model has converged and additional iterations do not lead to significant improvements in predictive performance.

The table below presents a comparative analysis of four machine learning algorithms—Lasso, Ridge Regression, Random Forest, and XGBoost—used to predict sea surface temperatures.

\begin{table}[h]
  \centering
  \caption{Model Performance}
  \begin{tabular}{lcccc}
    \toprule
    & Lasso & Ridge  & Random Forest  & XGBoost \\
    \midrule
    MSE & 0.2717821 & 0.2739555 & 0.01736646 & 0.04831045 \\
    MAE & 0.4145572 & 0.4167978& 0.28897168 & 0.17215728 \\
    RMSPE & 0.5213273& 0.5234076 & 0.41673080 & 0.21979639 \\
    KLD & 0.1316334 & 0.1319488 & 0.07246593 & 0.01819751 \\
    \bottomrule
  \end{tabular}
  \label{tab:example_table}
\end{table}

 The performance metrics evaluated include Mean Squared Error (MSE), Mean Absolute Error (MAE), Root Mean Squared Prediction Error (RMSPE), and Kullback-Leibler Divergence (KLD). Each metric provides a different perspective on the accuracy and reliability of the models.

The study finds that  Random Forest exhibits the lowest MSE (0.01736646), suggesting it has the highest accuracy in predicting sea surface temperatures among the four models.Also, XGBoost has the lowest MAE (0.17215728), indicating it provides more consistent predictions with smaller errors on average compared to the other models. XGBoost achieves the lowest RMSPE (0.21979639), highlighting its effectiveness in making accurate predictions. XGBoost has the lowest KLD (0.01819751), suggesting it provides the best fit to the actual data distribution.


\section{Conclusion}

Our results show that while Lasso and Ridge Regression offer competitive performance, advanced tree-based methods like Random Forest and XGBoost significantly outperform them in terms of all evaluated metrics. Random Forest identified Longitude (lon) and Latitude (lat), along with their interaction terms, as key predictors in sea surface temperature modeling. GCM variables (GCM.n1, GCM.n2, and GCM.n4) also emerged as crucial predictors, reflecting the influence of global climate patterns. XGBoost further highlighted the importance of General Circulation Models (GCMs) variables (GCM.n1, GCM.n2, GCM.n3, and GCM.n4) as the most significant features. Longitude (lon) and Latitude (lat) were also significant in the XGBoost model. While interaction terms contribute to the model, their impact is less significant compared to the Longitude, Latitude, and GCM components.

Random Forest achieves the lowest MSE, demonstrating its superior predictive accuracy. However, XGBoost consistently shows lower MAE, RMSPE, and KLD values, indicating it offers more reliable and precise predictions with fewer significant errors and a better approximation of the true data distribution. These results highlight the efficacy of using ensemble methods, particularly XGBoost, for predicting sea surface temperatures, making them valuable tools for climatological and environmental modeling.

\bibliographystyle{unsrt}  
\bibliography{references}

\end{document}